# On Adaptive Fairness in Software Systems

*Ali Farahani, *†Liliana Pasquale, ‡Amel Bennaceur, *Thomas Welsh, *‡Bashar Nuseibeh
*Lero – The Irish Software Research Center, Limerick, Ireland
†University College Dublin, Dublin, Ireland
‡The Open University, London, Uk
*name.lastname@lero.ie, †name.lastname@ucd.ie, ‡name.lastname@open.ac.uk

*Abstract*— Software systems are increasingly making decisions on behalf of humans, raising concerns about the fairness of such decisions. Such concerns are usually attributed to flaws in algorithmic design or biased data, but we argue that they are often the result of a lack of explicit specification of fairness requirements. However, such requirements are challenging to elicit, a problem exacerbated by increasingly dynamic environments in which software systems operate, as well as stakeholders' changing needs. Therefore, capturing all fairness requirements during the production of software is challenging, and is insufficient for addressing software changes post deployment. In this paper, we propose *adaptive fairness* as a means for maintaining the satisfaction of changing fairness requirements. We demonstrate how to combine requirements-driven and resource-driven adaptation in order to address variabilities in both fairness requirements and their associated resources. Using models for fairness requirements, resources, and their relations, we show how the approach can be used to provide systems owners and end-users with capabilities that reflect adaptive fairness behaviours at runtime. We demonstrate our approach using an example drawn from shopping experiences of citizens. We conclude with a discussion of open research challenges in the engineering of adaptive fairness in human-facing software systems.

## I. INTRODUCTION: FAIRNESS IN SOFTWARE SYSTEMS

Software systems are increasingly making decisions on behalf of humans, raising concerns about the fairness of such decisions. To date, researchers have focused on assessing fairness in such software systems by examining discrimination in different algorithms [1]–[6] and their implementations [2], [7], [8]. Much of the literature on fairness or discrimination in computer systems focuses on flaws in algorithmic design [2], [6]–[8] or biases in the input data of the system itself [4], [5], [9]–[12]. However, what is also notable is the absence of explicit and specific fairness requirements preceding algorithmic design, and the subsequent lack of checking or maintaining the satisfaction of those fairness requirements at runtime.

Discrimination attributed to input data can be managed by understanding the associated biases and mitigating their impact by defining new systems requirements [1], [13] or beforehand by using unbiased data for training the system [4]. For example, gender bias in a dataset can be addressed by explicit requirements for removing it from the data, or by designing the system to be sensitive to this bias [14]. Also, testing the input data or the software itself to find such biases, understanding their effects, and then fixing the software has been considered by researchers [9]–[11]. However, such approaches suffer the same issue: a lack of explicit fairness requirements at development time.

*Fairness Requirements* shape systems' behaviours to address stakeholders' perception of fairness. The elicitation and maintenance of such requirements is challenging, aggravated by such requirements being subjective, uncertain, and variable. Fairness is **subjective** and depends on the viewpoints of stakeholders as different stakeholders describe fairness requirements differently [2], [13], [15], [16]. As a result, they have different and sometimes contradictory requirements for one system [17]. For example, serving a protected group with priority can promote fairness in society, but at the same time, it may seem discriminatory to groups who do not benefit. Also, Fairness requirements are **uncertain,** and are often better understood once the system is deployed. Fairness requirements reflect different *policies* [1], [16] with different level of details [1], [13] and become more specific or complete over time. For example, awareness of gender bias in data may lead to the elaboration of existing equality policies to become more detailed in supporting gender equality [4]. Fairness requirements are **variable**. As such requirements are often described in terms of resources, and reasoning about these requirements involves reasoning about competition over shared resources [7]–[9], [11]. As a result, changes in resources, or stakeholders' perception of resources, can result in changes in the requirements for their fair distribution. For example, low stock of a resource may lead to additional restrictions on the maximum amount of the resource per user, in order to guarantee fair access of all users to the resource.

Fairness requirements may also reflect *multiple policies* about computations on shared resources [1], [7], [16], and therefore any effort to satisfy fairness requirements can affect the satisfaction of other requirements, due to different goals that stakeholders hold [8], [18]. Also, different stakeholders with different and perhaps conflicting goals may need to be addressed through trade-offs among these requirements [2], [13], [15], [19]. These two challenges about influences and conflicts between fairness requirements may become more problematic due to their changes post-development. There are existing approaches that support specifying and addressing conflicts in requirements, reasoning about their satisfaction, and adaptation of such requirements [20]–[25]. However, we suggest that challenges in satisfying fairness requirements can be better met by adopting more dynamic approaches to their monitoring and adaptation.

In this paper, we propose *adaptive fairness* as a means of *maintaining the satisfaction of changing fairness requirements in response to changes in the environment and consequently in resources*. We further suggest that adaptive fairness helps better understand fairness requirements, their relationship with resources, and different stakeholders' viewpoints. We characterise and model fairness requirements and discuss challenges to their satisfaction. We then explore how utilising models of **fairness requirements**, **resources**, and **operationalisation of fairness requirements** can achieve and maintain the satisfaction of fairness requirements. We suggest that these models can help integrate fairness requirements into existing approaches for requirements-driven adaptation. More generally, the paper contributes to the development of software systems that are more explicitly raise ethical concerns [26].

The rest of the paper is organised as follows. Section 2 reviews the literature on fairness in software systems. Section 3 presents a motivating example that illustrates the challenges of maintaining fairness in software systems in changing circumstances. Section 4 discusses fairness requirements and their satisfaction. Section 5 presents our approach to Adaptive Fairness. Finally, Section 6 discusses challenges and open problems.

## II. BACKGROUND AND RELATED WORK

We begin by reviewing multiple studies defining fairness in software systems from different perspectives. We also investigate previous attempts for addressing fairness in software systems and evaluate the possibilities of maintaining fairness in software despite frequent changes in fairness requirements.

### A. Defining and Addressing Fairness in Software Systems

There are different definitions of fairness in systems from different perspectives (e.g., machine learning [18], [19], [27], law and justice [2], [12], economy [28]). Dwork et al. [1] define fairness as "similar individuals should be treated similarly". This is aligned with the definition of fairness by Joseph et al. [6] who describes unfairness as preferentially choosing to serve an entity over another when the chosen one has lower expected quality (with respect to an important attribute). De Cremer et al. [29] and Folger et al. [30] treat "fairness" similarly to "justice". Fehr and Schmid [28] model fairness as "self-centred inequity aversion" where inequity aversion refers to users resisting inequitable outcomes of the system [8]. These definitions are used in the past for addressing fairness in software systems.

Achieving fairness in software systems, i.e. fairness requirements' satisfaction, is currently achieved by examining and later eliminating discrimination in the systems' outcomes [2], [11], [16], [27], [31]. The assessment of discrimination is performed by identifying and labelling agents as groups [32] or individuals [16], [32] and then examining if the system is discriminating against those groups or individuals. Another approach to assess discrimination is to inspect the system for the parity of the outcome based on the opportunity that an agent had and the effort that an agent exerted [28], [33]–[35]. However, defining fairness in software systems independent of the kind of discrimination, to which they lead, appears to be lacking. Also, the elimination of identified discrimination is suggested to take place either by reimplementing the system with the newly captured or emerged fairness requirements [19], [32] or tweaking the system at runtime by humans to satisfy those new requirements [12]. It is important to note that neither of these approaches are suitable for a system with dynamic fairness requirements in a dynamic environment. The dynamicity of the environment will cause frequent changes in fairness requirements, for example, because of changes in stakeholder's perception of fairness. Re-implementing the system or bringing humans in the loop for every change in fairness requirements is not practical or cost-effective. This suggests that an automated approach to address changes in fairness requirements is necessary.

### B. Automatically Satisfying Changing Fairness Requirements

Multiple papers that address fairness in software systems also propose a role for automation. The possibility of performing actions automatically to maintain or improve fairness in a software system has been criticised [12], raising issues about systems making decisions on behalf of humans. In contrast, Pessach and Shmueli [16] agree and review different techniques for automation of fairness. Other authors [5], [9], [11], [12] have proposed solutions for monitoring fairness in software systems by testing or assessing fairness criteria. Also, many authors [2], [15], [18], [19] have analysed trade-offs among stakeholders in their perceptions of fairness, and for the removal of discrimination from systems [8]. Selbst et al. [36] studied the elimination of discrimination in systems by learning the outcomes of actions on systems' fairness in order to plan for promoting fairness in systems. Yet, there is still a gap for addressing *changing* fairness requirements in the system in an automated manner.

There exists a considerable body of work [20], [21] on addressing changing software requirements as well as the adaptation of software requirements satisfaction. Also, related research has focused on addressing, predicting, and resolving conflicts in software requirements considering stakeholders and their viewpoints [22], [23]. Specifying and reasoning about the influences of the satisfaction of requirements on other requirements is another way of dealing with challenges in requirements-driven adaptation [24], [25]. Yet, to the extent of our knowledge, this large number of existing studies do not consider either the relationship between the requirements and resources or the stakeholders' viewpoints as important factors in their research. Bennaceur et al. considered software requirements tied to resources [37] and proposed a resource-driven requirements adaptation in response to the problems in the availability of resources. However, this approach does not tackle the issue related to the availability of resources and their impact on fairness requirements at the same time. As a result, we suggest that a new approach is therefore needed to maintain the satisfaction of changing fairness requirements that consider multiple stakeholders' viewpoints and the ties between fairness requirements and resources.

## III. MOTIVATING EXAMPLE

In this section, we present a motivating example drawn from shopping experiences of citizens. We consider different stakeholders, including *shoppers*, the *supermarket*, and the *government*. This is motivated by the Feed Me Feed Me exemplar and uses its terminology [38]. A shopping system of a supermarket supports searching items, adding/removing items to/from the shopping basket and later, checking out the basket. The following example shows that satisfying the initial fairness requirements is not sufficient due to subjectivity, variability or uncertainty. This leads to the emergence of new fairness requirements or changes in existing ones. Sometimes these requirements have conflicts due to differences in the goals of the stakeholders who specify them or resources that they address.

**Initial requirements:** This hypothetical system has been developed respecting some fairness requirements. An example of such requirements is: *shoppers should be treated equally for access to items (FR0).*

**A change in the environment:** We demonstrate our example in an environment with poorly predicted demands for items and unprecedented behaviours of shoppers (e.g., stockpiling), which are common during a pandemic. In our example, long waiting queues appear on the system, and accessing the online system becomes hard for some users with a bad internet connection or not being familiar with online shopping. Initial fairness requirements are not sufficient and a new requirement emerges: *elderly shoppers should have higher priority for accessing and using the online system during some periods (10 AM – 1 PM) during the working days (FR1)*. This requirement by the supermarket is aimed to achieve a fairer shopping experience, but it may affect other groups of shoppers and may have conflicts with *FR0*. However, dropping the *FR0* is not the best decision as it also covers situations that *FR1* does not cover (e.g., non-elderly shoppers). This is an example of the subjectivity of fairness requirements.

**A change in the resources:** Unpredicted changes can also happen in the availability of the resources (low stock for some items). This causes another requirement to emerge which will act as an extension to *FR0*: *shoppers should not buy more than 20 items per shopping trip (FR2)*. This is an example of the variability of fairness requirements as changes in resources cause the emergence of new requirements.

**An explicit need by a stakeholder:** As time passes and stakeholders gain enough knowledge about the environment, *FR2* seems insufficient. For example, the government proposes a new requirement mandating that the selling of health-related items should not be restricted, and all shoppers should be treated equally in this case. The proposed requirement is: *Health-related items should be sold without any restriction (FR3)*. This requirement is in conflict with both *FR2* and *FR1*. If shoppers have more than 20 items in his/her basket, *FR2* conflicts with *FR3* in this situation. This is an example of the uncertainty of fairness requirements which may lead to the emergence of new requirements due to a better understanding of the context.

## IV. Fairness Requirements and Their satisfaction

We define *Fairness Requirements* as requirements captured from different stakeholders to shape the system's behaviours in favour of the stakeholders' perception of fairness. In this section, we model fairness requirements and discuss the challenges related to maintaining their satisfaction at runtime.

### A. Defining Fairness Requirements

An important characteristic of fairness requirements is that they are **resource-oriented** [7]–[9], [11]. Any decision about whether it is fair to serve a group/individual or not is taken by using resources in the form of data from the environment [7]. Fairness requirements are also related to **multiple stakeholders** and can affect different groups and/or individuals [2], [13], [15], [16]. These are not only bounded to end-users but can include other stakeholders. In our example, *FR1* is required by the *supermarket* and will affect *shoppers* as well as the *government* (as it affects society). Fairness requirements can be operationalised by **multiple policies** aimed to control how the system performs computations on resources in a fair manner [1], [16]. For example, satisfying *FR1* can be achieved by the system either by blocking the access of shoppers younger than a certain age or modifying their basket and removing the item.

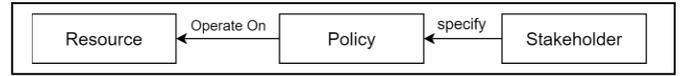

Fig. 1. Fairness Requirements Model

Consequently, we define **fairness requirements** as high-level policies specified by stakeholders that operate on resources (Fig. 1). These policies can have different levels of abstraction and detail [1], [13], which forms fairness requirements with different levels of abstraction too. However, similar to other software requirements, all of these fairness requirements within different levels of abstraction are contributing to the system's goals (in this case the fairness goals). Not all these requirements are refined enough to be satisfied and high-level or abstract fairness requirements need to be decomposed into *Operational Fairness Requirements*.

An operation fairness requirement is a fairness requirement in form of the tuple (s,p,r). This requirement is specified by a stakeholder ($s$) and is decomposed into policies ($p$) specified on resources ($r$) in the form of sets of rules and actions. A rule ($l$) is a set of conditions on resources and an action ($a$) is a description of computations on resources. For example, *FR1* is decomposed into two operational fairness requirements (*ofr1,ofr2*). The *ofr1* is *(s1,p1,{Shopper,Order})* and *ofr2* is *(s1,p2,{Shopper})*. In these two operational fairness requirements stakeholders (s1) are the *Government*, *Shoppers*, and the *Supermarket*. The *Supermarket* specified the requirements and the other two are affected by *ofr1* and *ofr2*. Even though the rules *l1* and *l2* within *p1* and *p2* are equal and *l1=l2={10<Date.time<13, Shopper.age<65}*, their associated actions are different. The action in *p1* is to remove the selected *Item* from the basket if the evaluation of the rules is true. In *p2*, the action is to redirect the *Shopper* out of the online shopping system. These operational fairness requirements are later used in section 5.

### B. Challenges of Fairness Requirements' Satisfaction

In this research, we consider that fairness requirements are precisely elicited and refined to operational fairness requirements. Adaptive Fairness tries to maintain the satisfaction of fairness requirements in the presence of changes in resources and considering challenges that happen because of a) conflicts in the satisfaction of fairness requirements or b) the influence of the satisfaction of fairness requirements on each other. Here, we will discuss these challenges in more detail.

Fairness requirements are subjective and may differ between stakeholders. These requirements can be conflicting due to the difference in stakeholders' goals [16] and knowledge of the environment. Therefore, trade-offs between these different requirements are necessary to overcome these conflicts [2], [16], [19]. Because of different knowledge of the environment, different stakeholders may pursue the same goal, but with different policies, each with different levels of details [1], [13]. Conflicts may be exacerbated if a policy addresses multiple resources or a resource is used in different policies. Therefore, the operationalisation of each fairness requirement can affect the resources that other fairness requirements depend on, and this may cause conflicts or problems in the satisfaction of fairness requirements.

Satisfying a fairness requirement impacts the satisfaction of other fairness requirements [8], [18] because of the relationship between fairness requirements and resources. Based on the presented structure for operational fairness requirements, the satisfaction of fairness requirements is achieved by performing computations on shared resources. Any computation on a resource can affect the result of other computations destructively or constructively considering the goal of their associated requirement [17], [18]. For example, the satisfaction of requirements concerned with limiting the number of *Items* that *Shoppers* can buy per shopping trip (*FR2*) can increase capacity for other *Shoppers*. If there is more shopping capacity there is no need for prioritising elders for shopping in a certain period or at least these periods can be shortened as all can be served without a problem.

Fairness requirements are uncertain which means that the list of elicitated fairness requirements may vary due to changes in fairness requirements or the emergence of new ones. This means the trade-offs among fairness requirements are also subject to change and more possible conflicts and unwanted influences may occur due to the variability and subjectivity of the requirements.

## V. On Adaptive Fairness

Adaptive Fairness is the notion of a customised requirements-driven adaptation for fairness requirements. This notion is about responding to situations where fairness requirements are evolving or emerging either because of changes in stakeholders' needs or resources. As shown in Fig. 2, the target systems' *Fairness Requirements* and *Resources* in the environment are the starting points of formulating the problem. By utilising models of **resources**, **fairness requirements,** and **fairness requirements operationalisation,** Adaptive Fairness offers to help the *Managing Element of an adaptive system* maintain the satisfaction of fairness requirements at runtime and overcome challenges around conflicts among such requirements (Fig. 2).

### A. Models for Adaptive Fairness

In the following section, models of fairness requirements, their operationalisation, and resources are described.

**Fairness Requirements Model (FR Model)** represents high-level fairness requirements and their decomposition into fairness requirements down to operational fairness requirements as leaf nodes. Also, this model holds the stakeholders for each requirement and these stakeholders can be refined in the process. Based on our earlier work on goal modelling [39], we use the goal modelling technique for representing FR Model. Capabilities of goal modeling in defining hierarchical goals, their decomposition and identifying/resolving possible conflicts in goals help the purpose of this research. This implies that as fairness requirements are software requirements, FR Model can be a part of a bigger requirements model, i.e. the system's goal model that represents the system's main objectives. This also helps using the approach in any existing requirements-driven approaches that perform over the system's goal model. Note that we use KAOS goal model notation [40] for the representation of the FR model.

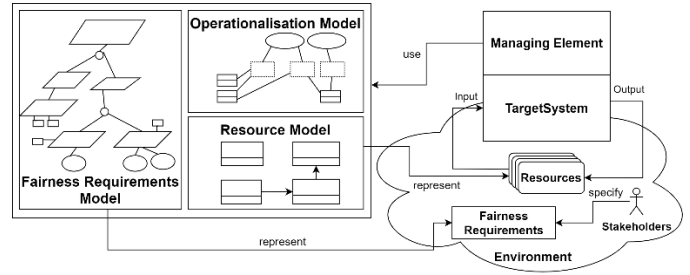

Fig. 2. Adaptive Fairness Approach

**Operationalisation Model** represents operational fairness requirements in the form of *operations* that have connections with resources. Each operation has a set of rules and actions and has connections to the resources. Operations are the same concept as *operations* in [40], however, here we use a simplified version of *operations* notation in KAOS. *Rules* are represented as conditions on resources and *Actions* are descriptions of computations on resources.

**Resources Model** represents different *resources* and their relationships. Resources are data entities representing actual elements of the environment. Hence, by relationships, we mean relationships of data objects such as association or composition. These representations can vary based on stakeholders as they perceive the environment with different details [13]. We represent resources by the notation of Entities in KAOS [40].

### B. Models in Actions

These three models make existing requirements-driven adaptation approaches capable of considering the relationships between fairness requirements and resources, as well as requirements and their stakeholders. Fig. 3 depicts the way that these three models are built and work together in action. The models are implemented for the motivating example. *FR0-FR3* are fairness requirements and they are decomposed to operational fairness requirements. These requirements are contributing to the more abstract fairness requirements (equal access to *Items*, prioritising protected groups) in the FR Model (Fig. 3). These requirements are defined by different stakeholders (*S1*, *S2*, and *G*). Also, some of the stakeholders (e.g., *S1*) are affected by some fairness requirements (e.g., *FR1, FR0, FR3*). *ORx-y* is an *operation* related to *OFRy* of the *FRx*. For simplicity in Fig. 3, we did not show the operational fairness requirements and only show operations and their connection directly to resources in Resource Model. We describe how the above models are meant to help maintain the satisfaction of fairness requirements in an adaptation feedback loop. For that purpose, we use the MAPE-K loop [41] as a reference model for the adaptation loop and discuss the contributions of Adaptive Fairness.

Our definition of fairness requirements provides information about stakeholders and their viewpoints. Also, the models are presented based on goal models. This makes Adaptive Fairness consistent with researches on resolving conflicts using viewpoints of stakeholders, such as [22], [23], or goal modelling techniques, such as [42]. To resolve conflicts in fairness requirements, Adaptive Fairness uses the presented three models to consider 1) stakeholders who are affected by fairness requirements and 2) fairness requirements ties with resources. Adaptive Fairness uses these ties as fingerprints for tracking possible influences among fairness requirements and use the stakeholders as a source for identifying conflicts between requirements. It also considers the conflicts in resources on which operational fairness requirements are operating, as a source of possible conflicts in fairness requirements. In the following, these possible extensions to the MAPE-K loop are exemplified.

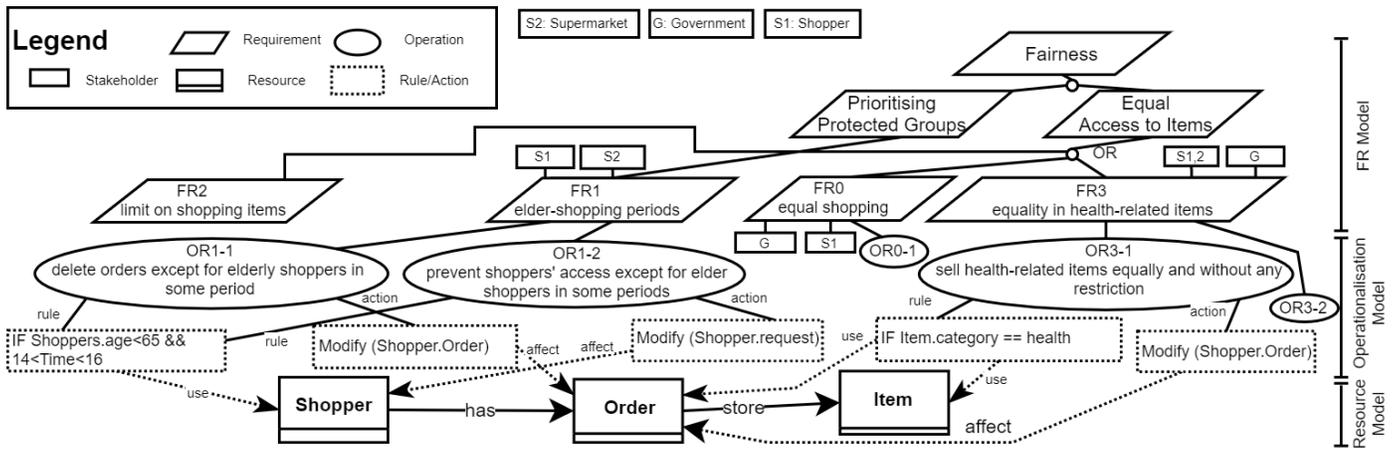

Fig. 3. Example of Requirements Model, Resources Model, and Operationalisation Model in Action

**Monitor**: This step monitors the environment to find any incident related to the unsatisfaction of any fairness requirements. For example, based on Operationalisation Model, the data about resources is gathered and used in evaluating rules inside the Operations. In our motivating example, let us consider that a *Shopper* with an age less than 65 is trying to add an *Item* to her/his basket at 3 PM. Also, let us consider that we are in the elder-shopping period. Thus, the result of the evaluation of *rules* inside *OR1-1*, *OR1-2*, and *OR1-3* is false (Fig. 3).

*Analyse*: In this step, the managing element will investigate whether any fairness requirements are unsatisfied based on the result of the Monitor step (in this case, evaluation of Rules). For example, if the result of the evaluation of rules inside *OR1-1, OR1-2*, and *OR3-1* is false, this can lead to unsatisfaction of *FR1* and *FR3 (Fig. 3)*. As mentioned before, unsatisfaction of *FR1* and *FR3* may happen due to their conflicts and therefore there is no need for a remediation plan.

Awareness about the used or affected resources by a fairness requirement as well as its stakeholders allows identifying possible conflicts between fairness requirements. Starting from each fairness requirement in FR Model and traversing over Operationalisation Model into Resources Model, all the resources that are in-use for the fairness requirement are extracted. The next step is to check other operations in Operationalisation Model to find operations that affect the same resources. The associated fairness requirements for these operations shows that their satisfaction can affect the first requirement. For example, as shown in Fig. 3, if *FR3* is unsatisfied and by checking its operations in the Operationalisation Model, we understand that *FR3* is sensitive to requirements affecting *Order.* Checking the Operationalisation Model shows that *FR1* affecting the same resource and unsatisfaction of *FR3* can happen because of an intentional satisfaction of *FR1*.

Also, using the information about the stakeholders of fairness requirements in FR Model can help resolve conflicts. This happens by checking the stakeholder who defines the requirement and the stakeholders who are affected. Examining the affected stakeholders of fairness requirements helps identify possible overlaps in the affected stakeholders of different requirements. These overlaps are signs of possible conflicts. For example, FR1 is specified by the supermarket and affects some *Shoppers*. FR3 is specified by the government and affects all the *Shoppers* (Fig. 3). As these two groups of affected stakeholders are overlapping, conflicts between the satisfaction of *FR1* and *FR3* is more likely to happen.

**Plan**: The task of this step is to prepare a plan out of possible options to satisfy the chosen fairness requirements. In this step, the focus is on building a plan for the satisfaction of a fairness requirement with less possible influences on other requirements.

Resources are the starting point for building a plan as they connect all fairness requirements in the system with each other through Operationalisation Model. First, based on the selected fairness requirements (in the previous step), actions are extracted from their associated Operations. Then the resources that they are using or affecting are extracted from Operationalisation Model. By examining alternative Operations, Adaptive Fairness helps to choose Operations that have fewer conflicts on resources. For example, let us consider that we are planning to satisfy *FR1* while *FR3* is already satisfied. There are two options for operationalisation (*OR1-1* and *OR1-2*). By choosing *OR1-1*, based on its rule, the resource *Order* is affected, and by choosing *OR1-2*, *Shopper* is affected (Fig. 3). Moving up from these resources toward Requirements (through the Operationalisation Model in Fig. 3) shows that the safest option is *OR1-2*. By choosing *OR1-1, Order* is getting affected and it is already in use by *OR3-1,* and this affects *FR3*. However, choosing *OR1-2* is not affecting any satisfied requirements.

If two operational fairness requirements are contributing to similar fairness requirements or goals (based on FR Model), most of the time they will follow the same strategy for tweaking with resources. The discussed calculation of conflicts does not consider the strategies behind each computation and therefore, its reported conflicts based on the resources can be ignored. But if two requirements are contributing to different requirements whom they or their parents are in an OR decomposition, the calculated conflict is more probable to happen due to the different strategies that their associated actions are following.

**Execute**: In this step, the actions for the satisfaction of fairness requirements are executed. The information about the decomposition of fairness requirements in FR Model and possible actions in Operationalisation Models can be used to order or prioritise actions. For example, let us consider that *FR1* and *FR2* are planned to be satisfied. The extracted actions in such a situation are 1) removing the *Order* of the *Shopper* and 2) preventing the *Shopper* from buying the *Item* by redirecting the request out of the system. It is clear that by applying the second action in the first place, the second action is undoable. However, by checking the parents of the requirements associated with *OR1-1* and *OR1-2* (*FR1* and *FR2*) this issue is avoidable. Implementing one of these requirements seems enough as they contribute to the same goal with an OR decomposition.

It should be mentioned that the three models are kept alive at runtime and serve as the Knowledge in the MAPE-K loop. Also, note that this research is about demonstrating the idea behind using these three models together for Adaptive Fairness. We neither discussed possible techniques for performing the adaptation nor the knowledge behind the adaptation. These are open problems that need special attention and we will address them in future work.

## VI. Challenges and Open Problems

We derived a number of open problems and challenges related to the elicitation of fairness requirements and reasoning about their satisfaction.

Although we presented the structural modelling of fairness requirements, developing a language for describing incomplete, evolving, and yet precise fairness requirements is still an open problem. Such a language should support the representation of *uncertainty of these requirements*, the *subjectivity of their goals,* and the *variability of their policies*. Further research is needed to integrate such requirements into the existing requirement engineering approaches. More precisely, further research is needed on refining these uncertain and subjective requirements into more operational ones at runtime, based on the stakeholder's needs and information from the environment.

The outcomes of satisfying fairness requirements in the long-term and short-term can be very different [17] and, therefore, they are *time-dependent*. For example, FR2 can promote fairness in the short term during the pandemic, but it will cause discrimination against other groups of people in the long term. The challenge is to predict the outcomes based on a time window and then plan for promoting fairness in the system. Such prediction is hard as the satisfaction of fairness requirements is achieved through computation on shared resources between different stakeholders. This means that the outcomes can unwillingly and unknowingly be different for different stakeholders at different times. This shows the importance of the *validity* and *verifiability* of such requirements, as well as the *explainability* of their satisfaction, as they can affect multiple stakeholders differently over time.

The satisfaction of fairness requirements influences other fairness requirements' satisfaction. Adaptive fairness attempts to address these influences. However, techniques for implementing such approaches may not be scalable. The number of possible influences between these requirements grows significantly as the number of fairness requirements increases. The number of these influences is related to the number of interconnections among different fairness requirements which grows significantly.

Fairness requirements' satisfaction can affect the satisfaction of other requirements (such as privacy, performance, usability and availability). *Privacy requirements* are about restricting access to personal data, but a system may need to access extra information about stakeholders or resources in order to satisfy fairness requirements. This raises challenges about whether or not satisfying fairness requirements interferes with satisfying privacy requirements. For example, information about the age of shoppers is needed to prioritise, say, older adult's access, but collecting and using such data adversely impact some users' privacy. Also, extra measures for computations on data can affect the *performance* of such systems in ways for which the systems were not designed. For example, checking the history of previous shopping transactions by each shopper at every turn can affect the overall performance of the system. Lastly, fairness requirements' satisfaction can affect the *usability* and *availability* of software for different stakeholders as it may pose restrictions to groups or individuals among the system's stakeholders.

## VII. Conclusion

In this paper we introduced adaptive fairness as a means for maintaining the satisfaction of changing fairness requirements in software systems, in response to changes in the environment and in resources. We investigated fairness requirements characteristics and proposed a model for representing such requirements. We described models of resources, fairness requirements, and their operationalisation, to help capture different viewpoints of stakeholders and the relationships between fairness requirements and resources. We demonstrated how these models can be built and how they contribute to overcoming challenges in maintaining fairness requirements' satisfaction. We are now extending our previous work [38] on resource-driven requirements adaptation to support the satisfaction of changing fairness requirements, and to suggest possible substitutions for resources that satisfice the stakeholders involved.

## Acknowledgements

This work was supported by SFI grants 13/RC/2094 and 16/SP/3804, and EPSRC grant EP/R013144/1.